\documentstyle[aps,prl,multicol,epsfig]{revtex}

\begin{document}

\title{Statistics of resonances and of delay times in quasiperiodic
Schr\"odinger equations}
\author{F. Steinbach, A. Ossipov, Tsampikos Kottos, and T. Geisel
\\
Max-Planck-Institut f\"ur Str\"omungsforschung und Fakult\"at Physik
der Universit\"at G\"ottingen,\\
Bunsenstra\ss e 10, D-37073 G\"ottingen, Germany}
\maketitle

\begin{abstract}
We study the statistical distributions of the resonance widths ${\cal P}
(\Gamma)$, and of delay times ${\cal P} (\tau)$ in one dimensional
quasi-periodic tight-binding systems with one open channel. Both quantities
are found to decay algebraically as $\Gamma^{-\alpha}$, and $\tau^{-\gamma}$
on small and large scales respectively. The exponents $\alpha$, and $\gamma$
are related to the fractal dimension $D_0^E$ of the spectrum of the closed
system as $\alpha=1+D_0^E$ and $\gamma=2-D_0^E$. Our results are verified
for the Harper model at the metal-insulator transition and for Fibonacci
lattices.
\\

\noindent PACS numbers:05.60.Gg, 03.65.Nk, 72.15.Rn
\end{abstract}

\begin{multicols}{2}

Quantum mechanical scattering, has been a subject of a rather intensive
research activity during the last years. This interest was motivated by
various areas of physics, ranging from nuclear \cite {nuclear}, atomic
\cite {atomic} and molecular \cite {molecular} physics, to mesoscopics
\cite {mesoscopics} and classical wave scattering \cite{S89}. The most
fundamental object characterising the process of quantum scattering is
the unitary $S$-matrix relating the amplitudes of incoming waves to the
amplitudes of outgoing waves. At present, there are two complementary
theoretical tools employed to calculate statistical properties of the
$S-$matrix, namely the semiclassical and the stochastic approach. The
starting point of the first is a representation of the $S-$matrix elements
in terms of a sum over classical orbits \cite{S89,KS99} while the later
exploits the similarity with ensembles of Random Matrices (see \cite{FS97}
and references therein). Thus, for chaotic/ballistic systems, many results
are known. Among the most interesting is the knowledge of the Wigner delay
time statistics and of the resonance width statistics \cite{FS97,BFB97}.
The former quantity captures the time-dependent aspects of quantum scattering.
It can be interpreted as the typical time interval a scattering particle
remains in the interaction region. It is related to the energy derivative
of the total phase shift $\Phi(E)$ of the scattering matrix i.e. $\tau(E)=
\frac{d\Phi(E)}{dE}$.  The resonances represent long-lived intermediate
states to which bound states of a closed system are converted due to coupling
to continua. On a formal level, resonances show up as poles of the scattering
matrix $S({\cal E})$ occurring at complex energies ${\cal E}_n = E_n - \frac
i2 \Gamma_n$, where $E_n$ and $\Gamma_n$ are called position and width of
the resonances, respectively.

Recently, the interest in quantum scattering has extended to systems showing
localization. For this case, there are analytical results about the
distribution
of phases of the $S-$matrix and of delay times
\cite{JGJ97,JVK89,TC99,OKG00,TB00}.
The former depends drastically on the disorder strength and energy
\cite{OKG00},
while for the latter a universal power law tail was found to hold \cite
{JVK89,TC99,OKG00,TB00}. Moreover, in \cite{TF00} a first analytical result
about the distribution of resonances appeared.

In the present paper we study delay time and resonance width statistics in a
new setting, namely a class of systems, whose closed system analogues have
fractal spectra. The latter exhibit energy level statistics that are in strong
contrast to the level repulsion predicted by Random Matrix Theory (RMT)
\cite{P65}. Their level spacing distribution follows inverse power
laws $P(s)\sim s^{-\beta}$ which is a signature of level clustering. The
power $\beta$ was found to be related with the fractal dimension of the
spectrum $D_0^E$ as $\beta = 1+ D_0^E$ \cite{GKP95}. Realizations of this class
are, quasi-periodic systems with metal-insulator transition at some critical
value of the on-site potential like the Harper model \cite{GKP95,AA80},
Fibonacci
chains \cite{GKP95,SBGC84}, or quantum systems with chaotic classical limit as
the Kicked Harper Model \cite{GKP91}. Here, for the first time we present
consequences of the fractal nature of the spectrum in open systems.
We consider open systems with one channel (the simplest possible
scattering problem) and report the appearance of a new type of resonances
width and delay time statistics. These distributions show inverse
power law behaviour dictated by the fractal dimension $D_0^E$ of
the spectrum. Specifically, we show that the probability distributions of
resonance widths ${\cal P} (\Gamma )$, and of delay times ${\cal P} (\tau )$
when generated over different energies, behave as
\begin{eqnarray}
\label{powlaw}
{\cal P}(\Gamma)&=&\Gamma^{-\alpha}\,;\,\,\,\alpha=1+D_0^E\nonumber\\
{\cal P}(\tau)&=&\tau^{-\gamma}\,;\,\,\,\gamma=2-D_0^E\,\,
\end{eqnarray}
For the calculation of ${\cal P}(\Gamma)$ and ${\cal P}(\tau)$ we employed
two independent approaches. Our results (\ref{powlaw}) are confirmed for two
different types of quasi-periodic tight-binding models and are supported by
analytical arguments.

We consider a 1D quasi-periodic sample of length $L$ with one semi-infinite
perfect lead attached on the left side. The system is described by the
tight-binding Hamiltonian:
\begin{equation}
\label {tight-binding}
H=\sum_n |n\rangle V_n\langle n| + \sum_n \left( |n\rangle \langle n+1| +
|n+1\rangle \langle n|\right)
\end{equation}
where $V_n$ is the potential at site $n$. In the sequel we will consider
examples where for $0\leq n\leq L$, $V_n$ is given by a quasi-periodic
sequence. For $n < 0$, $V_n=0$ and we impose Dirichlet boundary conditions
at the edge $\psi_{L+1} = 0$. Therefore, for $n\leq 0$, scattering states
of the form $\psi_n  = {\rm e}^{ikn} + S{\rm e}^{-ikn}$ represent the
superposition of an incoming and a reflected plane wave. Here, $k=\arccos(E/2)$
is the wave vector supported at the leads. Since there is only backscattering,
the scattering matrix $S(E)\equiv e^{i\Phi(E)}$ is of unit modulus and the
total information about the scattering is contained in the phase $\Phi(E)$.
One can write the scattering matrix in the form \cite{FS97,TF00,Fnote}
\begin{equation}
\label{smatrix}
S(E) \equiv  e^{i\Phi(E)} = 1-2i {w^2} \sin k\, \vec{e}^{\,T}
\frac 1{E-{\cal H}_{\rm eff}}\vec{e}.
\end{equation}
${\cal H}_{\rm eff}$ is an effective non-hermitian Hamiltonian given by
\begin{equation}
\label{Heff}
{\mathcal{H}}_{\rm eff}=H_L-w^2 e^{ik} \vec{e}\bigotimes\vec{e}.
\end{equation}
$H_L$ is the part of the tight-binding Hamiltonian (\ref{tight-binding})
with $n=0,....,L$ corresponding to the quasi-periodic sample and $\vec{e}=
(1,0,0,\ldots ,0)^{~T}$ is an $L-$dimensional vector that describes at which
site we couple the lead with our quasi-periodic sample. The strength of the
coupling is given by $w$. In the sequel we will always consider $w=1$.
Moreover,
since $\arccos (E/2)$ changes only slightly in the center of the band, we put
$E=0$ and neglect the energy dependence of ${\mathcal{H}}_{\rm eff}$. The poles
of the $S-$matrix are equal to the complex eigenvalues ${\mathcal E}$ of
${\mathcal{H}}_{\rm eff}$. The latter are computed by direct diagonalization
of ${\mathcal{H}}_{\rm eff}$. We note here that numerical diagonalization of
complex non-hermitian matrices is a time consuming process and imposes
limitations on the system size due to limited storage capacity. The
size of the matrices that we used in our analysis below was up to rank $5000$.

For the calculation of the Wigner delay time $\tau$ we have developed a simple
iteration relation in \cite{OKG00}
\begin{eqnarray}
\label{tau}
\tau_{L+1} &=& G_L^{-1} \left(\tau_L +\frac 1{sin k}\right) +
\frac {A_L \frac{cot k}{sin k}} {1+\left(tan(\phi_L-k)+A_L\right)^2}
 \nonumber\\
G_L &=& 1+A_L sin\left(2(\phi_L-k)\right)+A_L^2cos^2(\phi_L-k)\nonumber\\
\tan(\phi_{L+1}) &=& tan(\phi_L-k)+A_L
\end{eqnarray}
where $A_L=V_L/\sin k$. Iteration relation (\ref{tau}) has proved to be very
convenient for numerical calculations since it anticipates the numerical
differentiation which is a rather unstable operation. Moreover, it allows
us to go to large system sizes.

We motivate and numerically verify our results using first the well known
Harper model which is a paradigm of quasi-periodic 1D system with
metal-insulator
transition \cite{GKP95,AA80}. It is described by the tight-binding Hamiltonian
(\ref{tight-binding}) with on-site potential given by
\begin{equation}
\label{harper}
V_n=\lambda \cos (2\pi\sigma n ).
\end{equation}
This system effectively describes a particle in a two-dimensional periodic
potential in a uniform magnetic field with $\sigma=a^2eB/hc$ being the number
of flux quanta in a unit cell of area $a^2$. When $\sigma$ is an irrational
number the period of the effective potential $V_n$ is incommensurate with the
lattice period. We consider generic irrationals which cannot be approximated
``too well'' by rationals. To this end we take $\sigma$ as the limit of
successive rationals $p/q$, so that the potential becomes commensurate with
the lattice with period $q$. Then we can define a scaling procedure where
the incommensurate limit $q\rightarrow\infty$ becomes equivalent with the
thermodynamic limit. The states of the corresponding closed tight-binding
system are extended when $\lambda<2$, and the spectrum consists of bands
(ballistic regime). For $\lambda>2$ the spectrum is point-like and all
states are exponentially localized (localized regime). The most interesting
case is the critical point $\lambda=2$ where we have a metal-insulator
transition.
At this point, the spectrum is a zero measure Cantor set with fractal dimension
$D_0^E\leq 0.5$ \cite{frank} while the states are critical, i.e. self-similar
fluctuations of the wave function on all scales \cite{GKP95,AA80}.

First, we will investigate the statistical distribution of the resonance
widths $\Gamma$, and delay times for the Harper model at the critical point
$\lambda=2$. More exactly we determine the integrated distributions
\begin{equation}
\label{int1}
{\cal P}_{int}(x)= \int_{x}^{\infty} {\cal P}(x')dx'
\end{equation}
whose derivatives ${\cal P}(x) = -d{\cal P}_{int}/dx$ determine the
probability density of resonance widths ${\cal P}(x=\Gamma)$ and delay
times ${\cal P}(x=\tau)$. In all our calculations we will take approximants
of the golden mean $\sigma_G= ({\sqrt 5}-1)/2$. For this case it is known
that $D_0^E\approx 0.5$ \cite{frank}.

Figure 1 shows ${\cal P}_{int} (\Gamma)$ for two different rational
approximants $\sigma$ of the golden mean $\sigma_G$. It clearly displays an
inverse power law
\begin{equation}
\label{int1a}
{\cal P}_{int}(\Gamma) \sim \Gamma^{1-\alpha}
\end{equation}
and thus the resonance width distribution behaves as stated in (\ref{powlaw})
with $\alpha \simeq 1.5=1+D_0^E$. The integrated resonance width distribution
cuts off at a small value of $\Gamma$'s (see Fig.~1), since for all rational
approximants of $\sigma_G$ the total number of ${\cal E}_n$ is finite. This
cutoff, however, can be shifted to arbitrarily small values for higher
approximants.

Next we investigated the delay time statistics ${\cal P}(\tau)$. In Fig.~2 we
report the integrated ${\cal P}_{int}(\tau)$ for three different rational
approximants of the golden mean. Due to the efficiency of our iteration
relation
(\ref{tau}) we can approximate $\sigma_G$ by increasing the periodicity $q$ of
the potential as much as we like. Our numerical data are in agreement with an
inverse power law i.e.
\begin{equation}
\label{int1b}
{\cal P}_{int}(\tau) \sim \tau^{1-\gamma}
\end{equation}
with a value of $\gamma\approx 1.5 =2-D_0^E$ given by a best least square
fit, in perfect agreement with Eqn.~(\ref{powlaw}).

The connection between the exponents $\alpha, \gamma$ and the fractal dimension
$D_0^E$ of the close system calls for an argument for its explanation. The
following heuristic argument, similar in spirit to \cite{GKP95,BGS91} provides
some understanding of the power laws (\ref{powlaw}). We consider successive
rational approximants $\sigma_i=p_i/q_i$ of the continued fraction expansion of
$\sigma$. On a length scale $q_i$ the periodicity of the potential is not
manifest
and we may assume that a wave packet moves as $var(t)\sim t^{2D_0^E}$
\cite{roland}.
We attach the lead at the end of the segment $q_i$ which results in broadening
the energy levels by a width $\Gamma$. The maximum time needed for a particle
to
recognize the existence of the leads, is $\tau_{q_i}\sim q_i^{1/D_0^E}$. The
latter
is related to the minimum level width $\Gamma_{q_i} \sim 1/\tau_{q_i}$.
The number of states living in the interval is $\sim q_i$ and thus determines
the
number of states with resonance widths $\Gamma > 1/ \tau_{q_i}$. Thus ${\cal
P}_{int}
(\Gamma_{q_i})\sim q_i \sim \Gamma^{-D_0^E}$. By repeating the same argument
for
higher approximants $\sigma_{i+1}=p_{i+1}/q_{i+1}$ we conclude that ${\cal P}
(\Gamma)\sim \Gamma^{-(1+ D_0^E)}$, in agreement with (\ref{powlaw}). Although
the
numerical results support the validity of the above argument, a rigorous
mathematical proof is still lacking.

Next, we present another argument, which allows us to understand the relation
between
the power law decay exponent $\gamma$ and the fractal dimension $D_0^E$ i.e.
 $\gamma
=2 - D_0^E$.  Our starting point is the well known relation
\begin{equation}
\label{dtime}
\tau(E)=\sum_{n=1}^L \frac{\Gamma_n}{(E-E_n)^2 + \Gamma_n^2/4}
\end{equation}
which connects the Wigner delay times and the poles of the $S-$matrix. It is
evident
that anomalously large time delay $\tau(E)\sim \Gamma^{-1}_n$ corresponds to
the
cases when $E\simeq E_n$ and $\Gamma_n\ll 1$. In the neighbourhood of these
points,
$\tau(E)$ can be approximated by a single Lorentzian (\ref{dtime}). Sampling
the
energies $E$ with step $\Delta E\ll \Gamma_{min}$ we calculate the number of
points
for which the time delay is larger than some fixed value $\tau$. Assuming that
the
contribution of each Lorentzian is proportional to its width one can estimate
this
number as $\sum_{\Gamma_n < 1/\tau}\Gamma_n/\Delta E$. For the integrated
distribution
of delay times we obtain  ${\cal P}_{int} (\tau) \sim \int^{1/\tau} d\Gamma
{\cal P}
(\Gamma)\Gamma \sim \tau^{-(2 -\alpha)}$ in the limit $\Delta E \rightarrow 0$
where we used the small resonance width asymptotics given by
Eqn.~(\ref{powlaw})
(for similar argumentation see also \cite{FS97,DHM92}). Then for the asymptotic
distribution of delay times we get ${\cal P}(\tau) \sim \tau^{-(2 -D_0^E)}$ in
agreement with~(\ref{powlaw}) and our numerical findings.

The validity of the heuristic arguments (and thus of Eqs.~(\ref{powlaw})) can
be verified in more cases in the Fibonacci chain model of a one dimensional
quasi-crystal where other scaling exponents can be obtained. Here the potential
$V_n$ only takes the two values $+V$ and $-V$ arranged in a Fibonacci sequence
\cite{SBGC84}. It was shown that the spectrum is a Cantor set with zero
Lebesgue- measure for all $V>0$. We again find inverse power laws for the
integrated distributions ${\cal P}(\Gamma)$ and ${\cal P}(\tau)$. Here the
exponent depends on the potential strength $V$, while Eqs.~(\ref{powlaw}) still
relate the corresponding statistics to the fractal dimension $D_0^E$.  Our
results
for various $V$ values are summarized in Fig.~3 and show a nice agreement
between
the exponents $\alpha,\gamma$ and $D_0^E$ according to Eq.~(\ref{powlaw}).

Because of lack of space we defer the discussion of other results, like the
fractal nature of the resonance widths, and the behaviour of the delay time
autocorrelation function to a later publication \cite{OSKG00}.

We thank L. Hufnagel, R. Ketzmerick, H. Schanz, and M. Weiss for useful
discussions. (T.K) thanks U. Smilansky for initiating his interest in quantum
scattering.

\begin{figure}
\hspace*{-1cm}\epsfig{figure=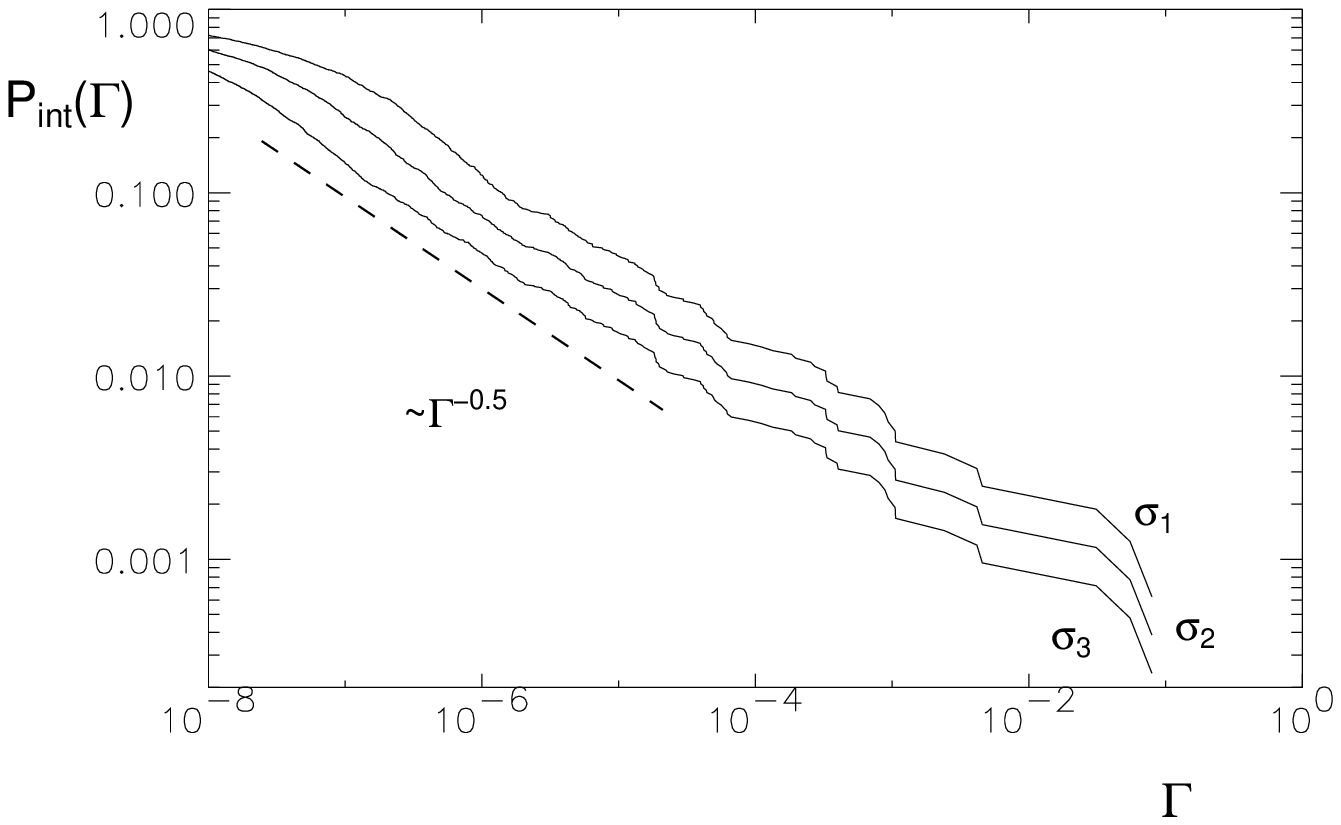,height=7cm,width=8cm,angle=0}
\noindent
{\footnotesize \\
{\bf FIG. 1.}
${\cal P}_{int}(\Gamma)$ of the Harper model $(\lambda=2)$
for three approximants of $\sigma_G$, $\sigma_1=\frac{987}{1597};
\sigma_2=\frac{1597}{2584};$ and $\sigma_3=\frac{2584}{4181}$.
An inverse power law $P_{int}(\Gamma) \sim \tau^{1-\alpha}$ is evident.
A least squares fit yields $\alpha
\approx 1.5$ in accordance with $D_0^E\simeq 0.5$ and Eqn.~(\ref{powlaw}).
As is seen the lower cutoff of the scaling region decreases for higher
approximants.}
\end{figure}

\begin{figure}
\hspace*{-1cm}\epsfig{figure=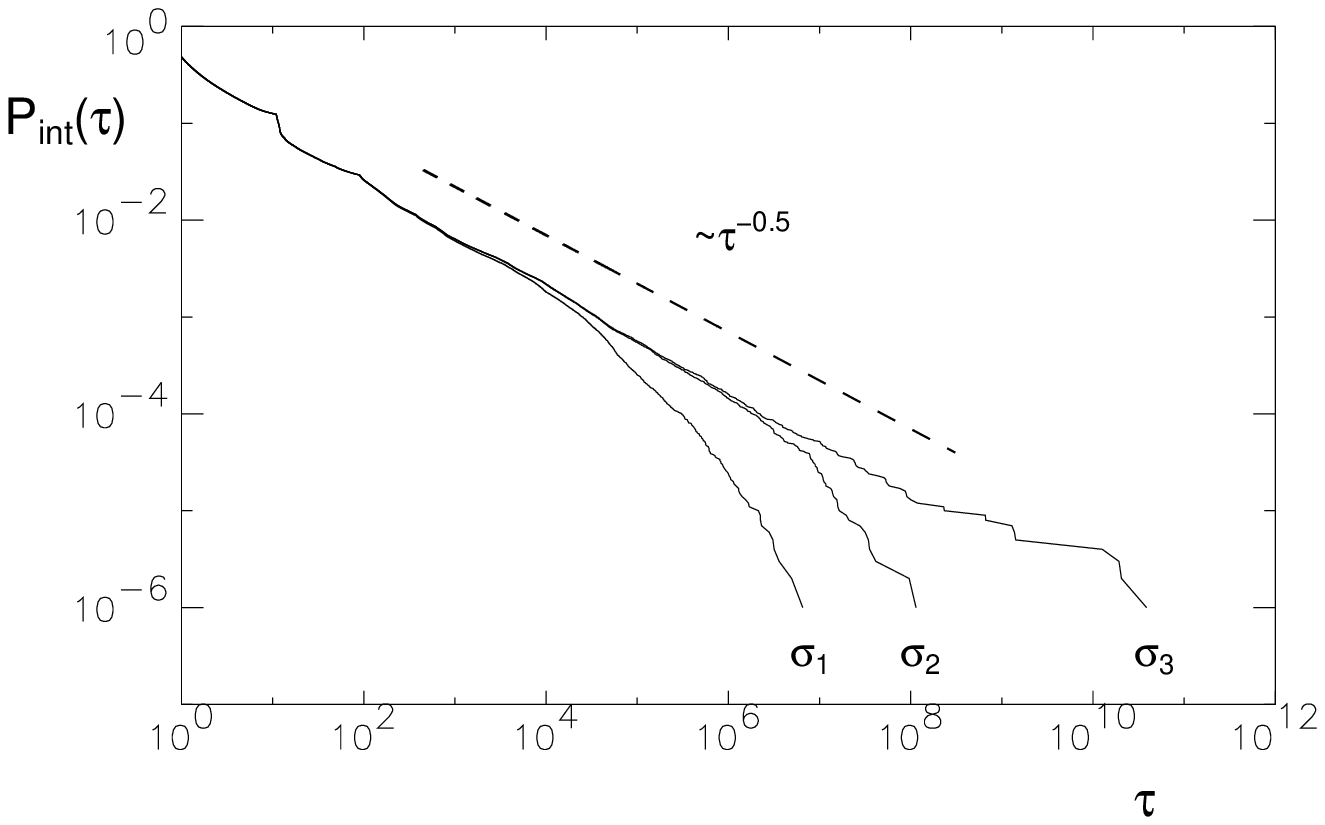,height=7cm,width=8cm,angle=0}
\noindent
{\footnotesize \\
{\bf FIG. 2.}
${\cal P}_{int}(\tau)$ of the Harper model $(\lambda=2)$ for three
approximants of the golden mean $\sigma_1=\frac{233}{377};
\sigma_2=\frac{987}{1597};$ and $\sigma_3=\frac{832040}{1346269}$.
An inverse power law $P_{int}(\tau) \sim \tau^{1-\gamma}$ is evident.
A least squares fit yields $\gamma\approx 1.5$ in accordance with
$D_0^E\simeq 0.5$ and Eqn.~(\ref{powlaw}). As is seen the upper
cutoff of the scaling region increases for higher approximants.}
\end{figure}

\begin{figure}
\hspace*{-1cm}\epsfig{figure=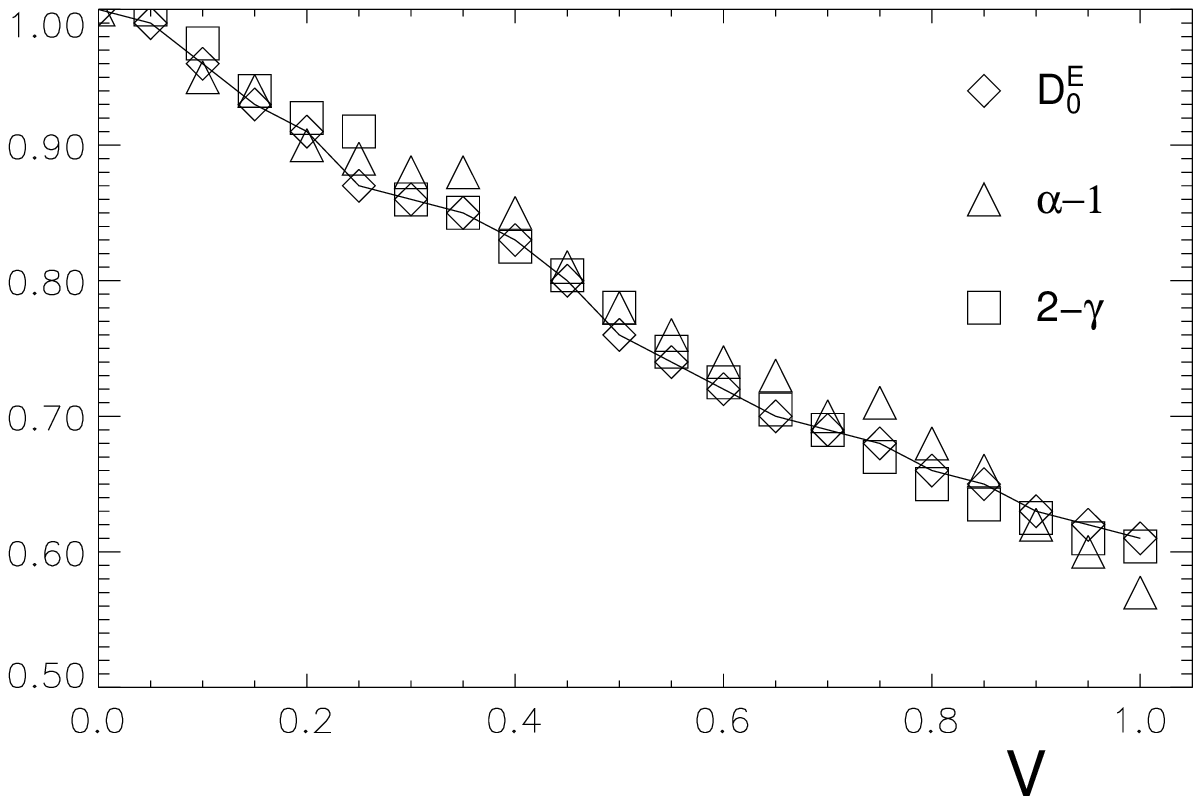,height=7cm,width=8cm,angle=0}
\noindent
{\footnotesize \\
{\bf FIG. 3.}
Power law exponents $\alpha,\gamma$ (plotted as $\alpha-1$ and $2-\gamma$)
of the resonance widths and of the delay time
distributions, respectively, as a function of the potential strength $V$ for
the Fibonacci model. We also plot the fractal dimension $D_0^E$ of the
spectrum (the solid line is to guide the eye). Our numerical data show
that $\alpha$ and $\gamma$ are related to the Hausdorff dimension $D_0^E$
according to Eqns.~(\ref{powlaw}).}
\end{figure}

\end{multicols}

\begin{thebibliography}{99}
\bibitem{roland} When averaged over different initial positions, the variance
of
a packet is believed to behave as $var(t)\sim t^{2D_{-1}}$ \cite{P99} instead
of $var(t)\sim t^{2D_0}$ as assumed here \cite{GKP95}. However, there is no
clear conclusion yet and one may consider our assumption on the spread of the
wavepacket as a good approximation. Our numerical results were not
conclusive at this point since $D_0\approx D_{-1}$.
\end{thebibliography}
\end{document}